\documentclass[
				aps,
				prl,
				reprint,
				superscriptaddress,
				showkeys
				]{revtex4-1}
\usepackage{graphics}
\usepackage{graphicx}				
\usepackage{epstopdf}

\usepackage{textcomp}
\usepackage{amsmath}

\usepackage{epstopdf}
\usepackage{color}

\begin{document}

	\title{Impact of electron solvation on ice structures at the molecular scale}
	\author{Cord Bertram}
	\affiliation{Physical Chemistry I, Ruhr-Universit\"at Bochum, D-44780 Bochum, Germany}
	\affiliation{Faculty of Physics, University of Duisburg-Essen, Lotharstr. 1, D-47057 Duisburg, Germany}
	\author{Philipp Auburger}
	\affiliation{Solid State Theory,
Friedrich-Alexander University Erlangen-N\"urnberg, Staudtstr.\ 7B2, D-91058 Erlangen, Germany}
	\author{Michel Bockstedte}
	\affiliation{Solid State Theory,
Friedrich-Alexander University Erlangen-N\"urnberg, Staudtstr.\ 7B2, D-91058 Erlangen, Germany}
	\affiliation{Chemistry and Physics of Materials, University of Salzburg, Jakob-Haringer-Str.\ 2a, A-5020 Salzburg, Austria}
	\author{Julia St\"ahler}
	\affiliation{Department of Physical Chemistry, Fritz Haber Institute of the Max Planck Society, Faradayweg 4-6, D-14195 Berlin, Germany}
	\affiliation{Department of Physics, Freie Universit\"at Berlin, Arnimallee 14, D-14195 Berlin, Germany}
	\author{Uwe Bovensiepen}
	\affiliation{Faculty of Physics, University of Duisburg-Essen, Lotharstr. 1, D-47057 Duisburg, Germany}
	\affiliation{Department of Physics, Freie Universit\"at Berlin, Arnimallee 14, D-14195 Berlin, Germany}
	\author{Karina Morgenstern}
	\affiliation{Physical Chemistry I, Ruhr-Universit\"at Bochum, D-44780 Bochum, Germany}

	\begin{abstract}

We determine the impact of electron solvation on D$_2$O structures adsorbed on Cu(111) with low temperature scanning tunneling microscopy, two-photon photoemission, and \emph{ab initio} theory. UV photons generating solvated electrons lead not only to transient, but also to permanent structural changes through the rearrangement of individual molecules. The persistent changes occur near sites with a high density of dangling OH groups that facilitate electron solvation. We conclude that energy dissipation during solvation triggers permanent molecular rearrangement via vibrational excitation.
	\end{abstract}
	
	\keywords{} 
	
	\maketitle


Excess electrons in polar environments interact strongly with their surrounding via the local Coulomb interaction, leading to electron solvation in polar liquids and amorphous solids \cite{sobolewski02,tauber2003,herbert01}, but also even to deformations of crystalline materials by small and large polaron formation \cite{onda01}. The build-up and subsequent stabilization of such species is driven by the minimization of the total energy of the system and coupled to nuclear motion in the hosting material \cite{emde01,tauber2003}. In water, the most ubiquitous solvent in nature, solvated electrons occur in the liquid and amorphous solid phase as well as in water clusters, and they localize at surfaces or in the bulk \cite{onda01,staehler01,bovensiepen01,herbert01}. While the structures of liquid water and ice are well-studied \cite{chandler01,sobolewski02}, experimental access to the solvent-solute complex structure is scarce and often relies on spatially averaging techniques \cite{malenkov01,young01}. As a consequence, our current understanding of the molecular scale details almost exclusively relies on theoretical models that, for instance, propose cavity and non-cavity pictures of electron solvation \cite{chandler01,discussion} or encompass various idealized model structures and locally varying environments at ice surfaces \cite{chandler01}. Notably, the common feature of such models ranging from surface voids to orientational defects is a local agglomeration of dangling OH groups.\par 
Beyond the fundamental need of molecular scale insight into the local, equilibrated environment of solvated electrons, detailed understanding of its formation process and interaction with the surrounding molecules is equally crucial: The energetic relaxation of excess electrons in aqueous solution is known to be connected to transient molecular rearrangements \cite{herbert01,onda01} that begin on femtosecond timescales. This nuclear motion is triggered by the localized charge and reversed in its absence when the system relaxes towards its neutral ground state. However, solvated electrons are suspected to also induce \emph{permanent} changes to water structures \cite{chakarov01, king01}. For example, crystallization of amorphous solid water \cite{chakarov01} supported by a solid template under UV light illumination was interpreted as a consequence of energy transfer during electron trapping. Such fundamental modifications of hydrogen-bonded water structures require remarkably strong interactions between the solvated electron and its environment. This demands local, molecular scale probes that are capable of the identification and characterization of active sites before and after their population. Femtosecond time-resolved two-photon photoelectron (2PPE) spectroscopy experiments showed that electrons in conductive templates are excited by UV photons and either transferred to preexisting electron traps or the conduction band of adsorbed ice structures \cite{bovensiepen01,staehler03,onda01}. Combined with scanning tunneling microscopy (STM), a direct link between the solvated electron dynamics and ice morphology was established \cite{staehler01}, however, molecular scale insight could not be achieved so far.\par
In this Letter, we demonstrate that solvated electrons
do not only induce transient, but also permanent rearrangements
of water molecules 
on ice surfaces. Shown locally by STM and confirmed by \emph{ab initio} calculations, these rearrangements originate from
photoexcited electron solvation. This first direct observation
of significant molecular scale modifications of ice structures
caused by solvated electrons highlights
a concerted action of dangling OH-groups in electron localization.\par
The measurements are performed in two separate ultra-high vacuum chambers equipped with well-established facilities for surface preparation and characterization. 
The Cu(111) surface is cleaned by standard sputtering and annealing cycles (for details \cite{SI}).
Amorphous ice is deposited on Cu(111) held at 88~K with a flux of 0.1~BL/min for 9~min for a coverage of 0.9~bilayers (BL) \cite{endnote}.  
The ice is crystallized by annealing for 12~minutes at 114~K.\par
For the STM measurements, performed below 11~K, the system is excited by a tunable laser providing wavelengths between 330~nm and 450~nm (3.8~eV and 2.8~eV). 
At these temperatures, thermal motion or reorientation of water molecules does not occur. As the UV photons are not absorbed in the ice structures, all structural rearrangements are driven by electron-induced relaxation processes.
At a grazing incidence angle of 79$^\circ$, the wavelength-dependent absorbed fluence varies between 7.0~pJ/cm$^2$ and 36.2~pJ/cm$^2$ in the region directly below the tip. 
Different irradiation times between 3 h and 16~h are used to compensate for differences in fluence. Before irradiation, the tip is moved at least 500~nm away from the laser path and from the measured region in order to avoid near-field effects (for details cf.\ \cite{SI}).\par
	For 2PPE, performed at 30~K, a tunable regeneratively amplified femtosecond laser system provides laser pulses at h$\nu_\mathrm{pump}$=3.99~eV and h$\nu_\mathrm{probe}$=2.04~eV to photoexcite and photodetach the solvated electrons, respectively \cite{bovensiepen01}. Their kinetic energy is measured by a time-of-flight spectrometer as a function of the variable time delay $\Delta t$ between ${h\nu_\mathrm{pump}}$ and ${h\nu_\mathrm{probe}}$ and is referenced to the substrate's Fermi energy $E_\mathrm{F}$ (see \cite{SI} for details). \par
 
DFT with gradient corrected (PBE) functional is used for structural optimization. Electronic properties are described within Hybrid DFT (PBE0) and many-body perturbation theory using the GW approximation (for details \cite{SI}). By \emph{ab initio} modeling, we address the electron affinity of selected admolecule structures.\par
In the following, we first conclude on the electron solvation probability in amorphous as compared to crystalline ice. Subsequently, we identify molecular structures which exhibit permanent changes induced by electron solvation. We finally proof the effect of electron solvation via a systematic photon energy dependence.\par 
We start by reproducing and characterizing the ice structures of interest here \cite{mehlhorn02}.
	\begin{figure}[ht!]
		\includegraphics[width=0.99\columnwidth]{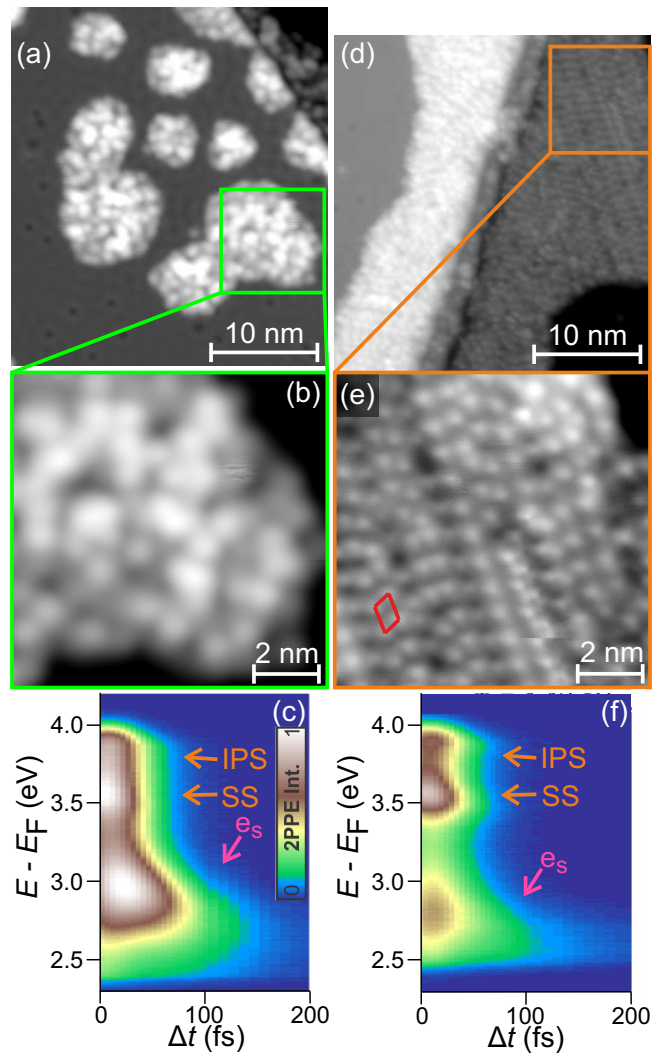}
		\caption{Ice structures and dynamics: 
		(a-c) amorphous porous ice (d-f) crystalline ice (a,b,d,e) STM images, red rhomboid in (e): unit cell of ($2\times1$) superstructure on top of a hexagon adrow (a,b: V = 0.1~V, I = 8~pA; d,e: V = -0.1~V, I = 2~pA) (c,f) 2PPE intensity in false colors as a function of time delay $\Delta t$ and of energy above Fermi energy ${E-E_\mathrm{F}}$; spectra recorded consecutively under identical experimental conditions (c) data reproduced from Ref.\ \cite{staehler01}; e$_\mathrm{s}$: transiently solvated electrons; IPS: image potential state; SS: surface state of Cu.}
		\label{fig_01}
	\end{figure}
At the deposition temperature of 88~K used here, the amorphous porous ice decorates the step edges on Cu(111) and forms islands on the terraces with diameters between 3~nm and 10~nm (Fig.\ \ref{fig_01}a) \cite{mehlhorn02}. 
Its amorphous nature is most obvious in the distribution of protrusions of different sizes on top of the islands without any 
order (Fig.\ \ref{fig_01}b).\par
After crystallization, the resulting structure consists of hexagonal ice bilayers \cite{mehlhorn02}. The number of islands on the terraces is reduced, their width is increased up to 30~nm, and the step edges are completely covered by ice (Fig.\ \ref{fig_01}d). 
 The crystalline nature is most obvious in the long range order of single water molecules on top of
the islands, each of them imaged as a protrusion of uniform size (Fig.\ \ref{fig_01}e, \cite{bockstedte01}). 
 Previous studies showed that these molecules reside on top of two complete and one to three partially filled BLs \cite{mehlhorn02}. The partial BLs form different types of local superstructures, e.g., ($2\times2$) or ($2\times1$) (see unit cell in Fig.\ \ref{fig_01}e, \cite{mehlhorn02}). All partially filled bilayers contain molecules that are bound to only two or three other water molecules \cite{bockstedte01}, in contrast to the water molecules in bulk ice with four binding partners \cite{thiel01}. Such a reduced number of binding partners leads locally to a substantial number of additional polar dangling OH groups.\par 
Time-resolved 2PPE spectra of amorphous and crystalline ice demonstrate how the structures differ with respect to electron solvation (Fig.\ 1c,f). For the porous amorphous ice structures, the electrons initially exhibit an intermediate state energy of ${E-E_\mathrm{F} = 2.97}$~eV (Fig.\ \ref{fig_01}c) and stabilize in energy with a rate of 220~meV/ps \cite{staehlerxx}. 
Here, we compare these results to the solvation of electrons on the crystalline ice islands after annealing (Fig.\ \ref{fig_01}f). All spectral signatures are identified for both structures, but the intensity of the solvated electron signature $e_S$ is significantly reduced for the crystalline ice.\par
\begin{figure}[ht!]
		\includegraphics[width=0.99\columnwidth]{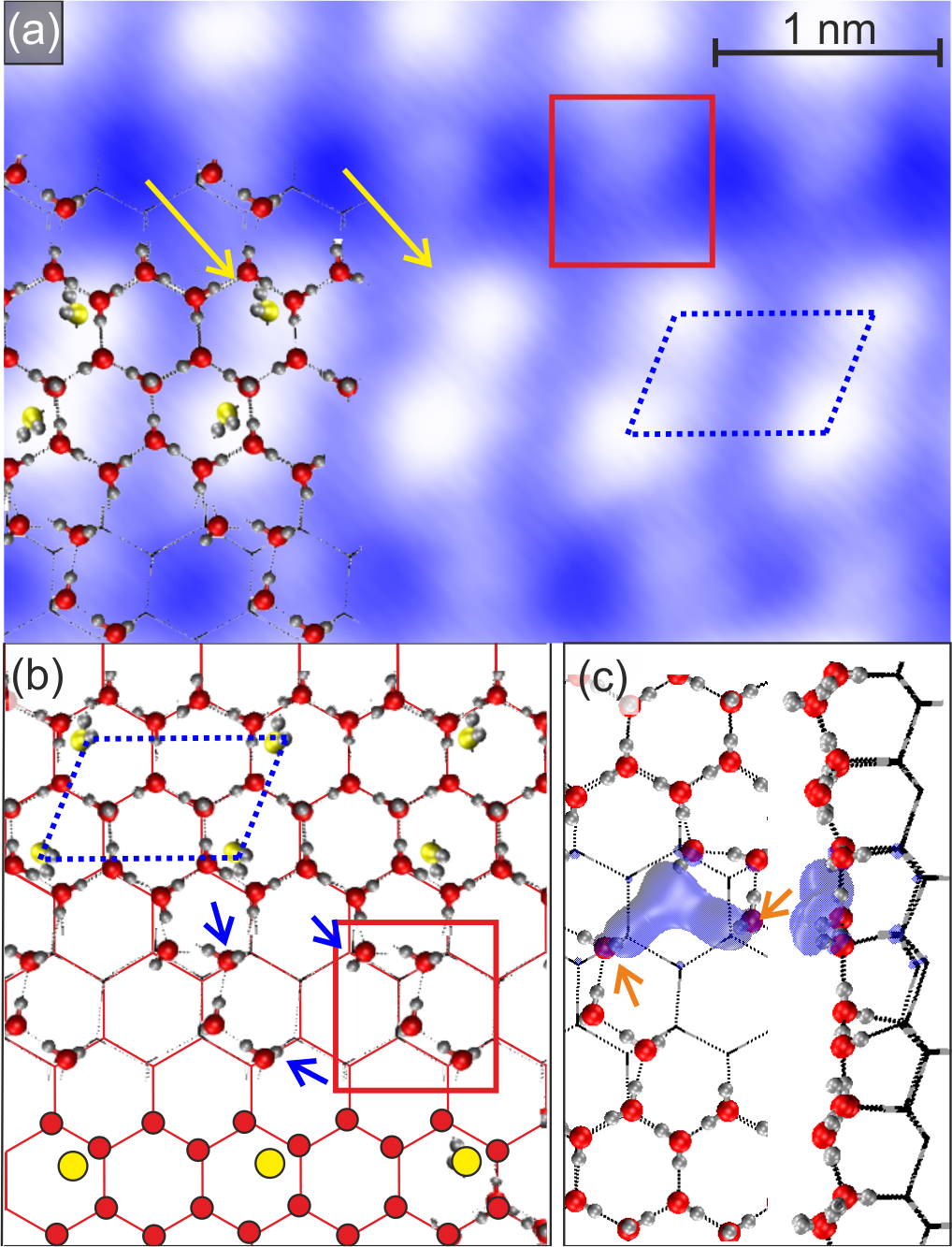}
		\caption{Structure model for electron solvation: (a) high resolution STM image of crystalline ice with ($2\times1$) superstructure (dashed blue rhomboid) and interconnecting tetramers (red rectangle); arrows point to individual molecules in superstructure (-0.1~V, 2~pA) (b) scheme of structure based on \cite{bockstedte01}; red hexagonal lattice represents upper complete BL; molecules in first and second partial BLs with oxygen atoms in red and yellow, respectively; hydrogen atoms in white; arrows point to some dangling OH groups; in lower left corner molecules represented by circles as used in Fig.\ \ref{fig_03a} (c) top view (left) and side view (right) of calculated solvated electron density of solvated electron between interconnecting tetramers, blue surface represents iso-value of the charge density of 12~nm$^{-3}$; arrows point to two reoriented molecules.}
		\label{fig_02}
	\end{figure}
	To relate this drop in intensity to the geometric difference between the structures, we aim at identifying preferred solvation sites. We exemplify this identification for a frequently observed crystalline structure, the ($2\times1$) superstructure (Fig.\ \ref{fig_02}). Individual molecules within the superstructure, marked by water molecules with oxygen atoms colored in yellow, reside on top of a partial BL. The partial BL is formed by rows of two hexagons in width 
	connected by chains of different lengths, here a tetramer (red rectangle). 
	Note that each of the molecules within the tetramer is slightly displaced from the exact on-top adsorption site in a complete BL. 
Our calculations confirm that the dangling OH groups in the interconnecting tetramers (blue arrows) promote electron solvation. For instance within a hexamer between hexagon adrows, the electron localizes between three dangling OH groups, from which two reorient towards the excess electron  (Fig.\ \ref{fig_02}c, orange arrows). This reorientation provides a binding energy gain of 570~meV compared to the lower edge of the ice conduction band, well within the previously observed range for electron localization in ice structures of up to 0.6~eV \cite{bovensiepen01}. Other arrangements of admolecules between the hexagon adrows identified in STM
images and calculated by DFT (see also \cite{SI}), such as
dimers and trimers, exhibiting one to two neighboring
dangling OH groups, yielded lower binding energy gains
of 100 meV and below. Importantly, for other tested orientations of
water molecules in the tetramer with fewer dangling OH
groups no electron solvation was found.
Our calculations of admolecule structures on crystalline ice thus show that an agglomeration of dangling OH groups embedded in a polar hydrogen bond network promotes electron solvation. Comparable sites are, however, much more abundant in amorphous structures, explaining the larger number of solvated electrons measured by 2PPE (Fig.\ 1c).\par 
\begin{figure}[ht!]
		\includegraphics[width=0.99\columnwidth]{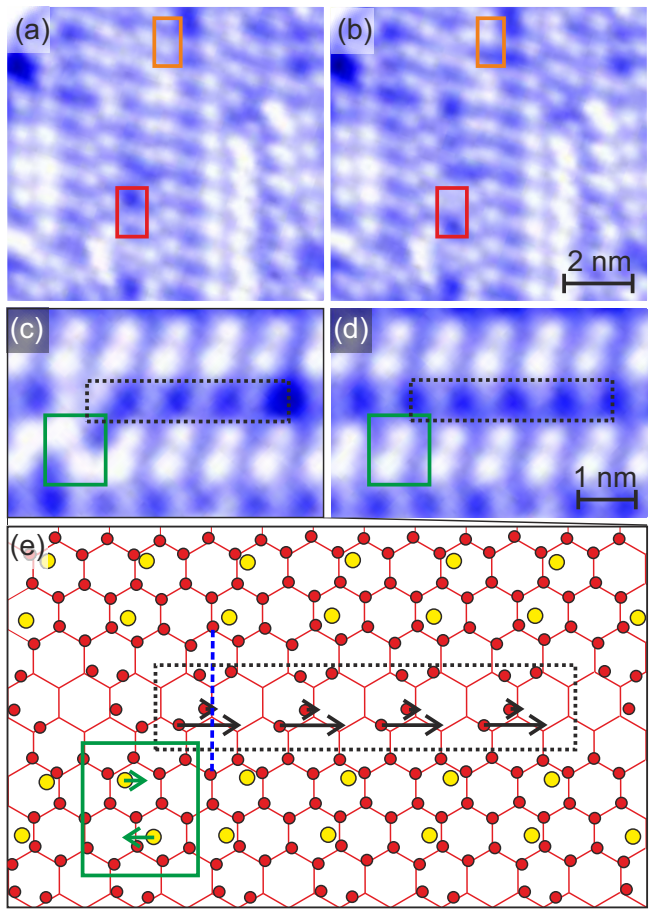}
		\caption{Photoinduced changes to crystalline ice structures: (a-d) STM images (a,c) before and (b,d) after irradiation (a,b) for 9.5~h with 330~nm photons (3.77~eV, $6.9\cdot10^{4}$~photons/molecule) and (c,d) for 15 h 11 min with 400 nm photons (3.11~eV, $2.6\cdot10^{5}$ photons/molecule), some changes indicated by rectangles (-0.1~V, 2~pA) (e) scheme of changes from (c) to (d); green and black arrows indicate direction and length of rearrangement; color code as in Fig.\ \ref{fig_02}b.}
		\label{fig_03a}
	\end{figure} 
After having discussed the importance of dangling OH groups for electron solvation on the molecular level, we now describe the effects of illuminating crystalline ice structures by UV light. 
In the example shown, orientational defects and vacancies interrupt the long range order of the superstructures (cf.\ rectangles in Fig.\ \ref{fig_03a}a and c). Furthermore, the orientation of four interconnecting tetramers is turned by roughly 30$^{\circ}$ clockwise as compared to the interconnecting tetramers of the surrounding superstructure (dashed, black rectangle). 
Changes to such structures are observed during irradiation at different wavelengths below 11 K, i.e.\ far below the temperature for hydrogen bond rearrangements of $\approx100$~K. 
For example, a single water molecule on top of the adrow moves in direction of a vacancy (upper rectangle, Fig.~\ref{fig_03a}a to b). One interconnecting tetramer moves into the neighboring vacancy thereby creating a vacancy at its previous position (lower rectangle, Fig.~\ref{fig_03a}a to b). 
In the second example, all molecules match the long-range periodicity (Fig.\ \ref{fig_03a}c to d).
We observe such permanent rearrangements close to structural defects.

By comparing the high resolution images with our \emph{ab initio} calculations, we develop a molecular-scale picture for the two rearrangements shown in Fig.~\ref{fig_03a}c,d. 
Concerning the defect in the ($2\times1$) superstructure, the molecules move to a neighboring adsorption site (Fig.\ \ref{fig_03a}e, green arrows). Since the involved water molecules are bound by two hydrogen bonds, vaulting one molecule over another one requires the rearrangement of one hydrogen bond only. 
\par  
The change in the interconnecting tetramers corresponds to a mirroring of each interconnecting tetramer (dashed blue line in Fig.\ \ref{fig_03a}e). While one molecule moves to a next neighboring adsorption site (longer black arrow), the other one simply adopts its energetically most favorable site to the new binding situation shifting by less than 0.1~nm (short black arrow). The rearrangement of the whole tetramer is thus initiated by the solvated electron induced rearrangement of a single molecule.\par 
Such structural rearrangements could only be revealed because we are able to image exactly the same spot of the sample before and after its illumination. They can only be understood by the aid of state-of-the art \emph{ab initio} calculations. STM is, however, much too slow to follow the short-time dynamics of the solvated electrons which demands time-resolved 2PPE. Only time-resolved 2PPE shows the electron dynamics and energies involved, necessary to understand the physical origin of the permanent rearrangements. Because of its wide band gap, direct photoexcitation of ice would require three photons in our experiments. As three photon processes are rather unlikely, the photons are predominantly absorbed in the metal substrate. Any energy necessary for molecular rearrangement must thus be transferred from there to the water molecules. The solvated electrons dissipate up to 0.6~eV as measured by 2PPE \cite{bovensiepen01}, consistent with the calculated values. This large amount of energy gained by electron localization and stabilization is released during solvation and well-capable of inducing permanent rearrangements.  Theory revealed that solvation demands the concerted action of dangling OH groups and STM showed that permanent rearrangements happens close to defects, which theory describes as a source of dangling OH groups. We conclude that the permanent rearrangements are initiated by solvated electrons.\par 
\begin{figure}[ht!]
		\includegraphics[width=0.99\columnwidth]{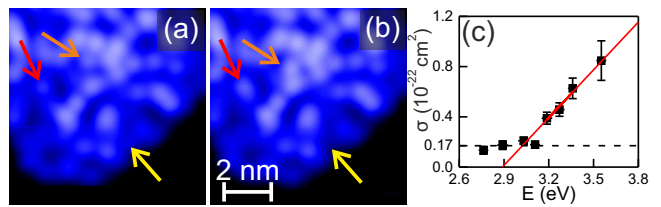}
		\caption{Photoinduced changes to amorphous porous ice: (a,b) STM image (a) before and (b) after irradiation with 390 nm for 7~h (3.19~eV, ${\sim2.5\cdot10^{5}}$ photons/molecule); some of the changes indicated by arrows (0.1~V, 8~pA) (c) cross section of changes $\sigma$ versus photon energy $E$, black line: mean below threshold; red line: linear fit to rising region.}
		\label{fig_04}
\end{figure}
We finally quantify the impact of electron solvation. The number of photoinduced rearrangements should increase for photon energies that can lift electrons from the metal into the ice, which exhibits unoccupied states available for $h\nu>3.0$~eV as determined by 2PPE (cf.\ Fig.\ \ref{fig_01}c). For better statistics, we determine the number of photoinduced rearrangements $N_E$ on amorphous ice with its larger density of dangling OH groups, as exemplified by Fig.\ \ref{fig_04}a and b. Also for the amorphous structures, we can follow the motion of individual molecules and clusters during irradiation in our specific set-up. A substantial number moves into different directions, either parallel to the ice surface (orange arrow) or perpendicular to it (red and yellow arrows). As in the case of admolecule rearrangement on crystalline ice (see Fig. \ref{fig_03a}), the shape of the protrusions is preserved during rearrangement ensuring intact molecules. From these STM observations, we determine the cross section $\sigma=\frac{N_E}{A_{island}}\cdot\frac{1}{\rho_{ph}\cdot\rho_{D2O}}$ with $A_{island}$ representing the fraction of the surface area covered by ice, the density of absorbed photons $\rho_{ph}$, and the density of water molecules per surface area $\rho_{D2O}$ for photon energies between 2.70 and 3.65 eV (for details see \cite{SI}). The cross section describes the probability of a photoinduced molecular rearrangement.\par 
As shown in Fig. 4c, the cross section remains almost constant at $0.17\cdot10^{-22}$~cm$^2$ for photon energies up to $(3.0\pm0.1)$~eV and then increases substantially with $(1.3\pm0.2)\cdot10^{-22}$~cm$^2$/eV. The threshold corresponds to the minimum energy needed to inject electrons into the ice conduction band at 3.0~eV as determined by 2PPE above. Furthermore, the increase up to 0.7~eV above the threshold agrees well with an rise of the population of solvated electrons on crystalline ice on Ru(100) in a similar energy range \cite{bovensiepen01}. Threshold and increase corroborate that solvated electrons are responsible for the observed permanent changes to the ice structures. With an estimated $10^{-2}$ solvated electrons per incident photon in our work (see \cite{SI}), the range of determined cross sections translates to $10^{-5}$ rearrangements per excess electron. This is in good agreement with typical values for electron-induced rearrangements in inelastic electron tunneling experiments \cite{morgenstern,mehlhorn2009}.\par 
How can solvated electrons lead to permanent rearrangements of molecules? A rearrangement requires breaking of at least one hydrogen bond, which costs around 140~meV in the ($2\times1$) superstructure or 163~meV for the interconnecting tetramer \cite{bockstedte01}. 
We propose the following scenario:\\
$\bullet$ First, the photoexcited electron reaches the bottom edge of the ice conduction band.\\ 
$\bullet$ Subsequently, it localizes near dangling OH groups, a process that releases $>420$~meV within only 20~fs, followed by electron solvation with an energy dissipation rate of 220~meV/ps \cite{staehlerxx}. In insulating materials like ice, the only energy dissipation channel within this energy range is the vibrational excitation of molecules surrounding the solvated electron. As the energy has to be dissipated within less than a ps after electron injection for both, crystalline and amorphous ice structures \cite{staehler03}, we expect the energy to be initially distributed among a few directly neighboring molecules only.\\
$\bullet$ Prior to dissipation to other molecules, the vibrational excitation of higher energy modes can be redistributed via anharmonic coupling to modes relevant to molecular rearrangement between different isoenergetic ice polymorphs \cite{komeda01}. 
Excitation of frustrated translational and frustrated rotational modes results eventually in the observed molecular rearrangement. For this process it is irrelevant, whether  the electron decays back to the metal surface before or after the permanent rearrangement of the water molecules.\par 
In conclusion, our combined study reveals the impact of solvated electrons at the surface of condensed ice phases. We traced back the local, permanent rearrangements to electron solvation facilitated by a concerted action of dangling OH groups. This process is not only expected for ice supported by other metals, but for all ice structures, which show electron solvation with major energy dissipation. Possibly, the vibrations induced by this energy dissipation could be detected on the local scale with future experimental techniques, see \cite{cocer00}. 

	\begin{acknowledgments}
		The authors acknowledge financial support by the Deutsche Forschungsgemeinschaft under grant MO960/18-1, BO1841/3-1, SFB 450, and through the Cluster of Excellence RESOLV (ECX 2033). Computing time was granted by the Research Center J\"ulich (HER140) and on the HPC-cluster of the RRZE of FAU Erlangen-N\"urnberg.  
	\end{acknowledgments}



\begin{thebibliography}{9}
\bibitem{endnote} A bilayer is a layer of water molecules in a hexagonal arrangement with the molecules in two slightly displaced layers and used as a mass equivalent \cite{thiel01}.
\end{thebibliography}
\end{document}